\documentclass[10pt,conference]{IEEEtran}
\IEEEoverridecommandlockouts
\usepackage{silence}
\usepackage{cite}
\usepackage{amsmath,amssymb,amsfonts}
\usepackage{graphicx}
\usepackage{xcolor}
\usepackage{float}
\usepackage{subcaption}
\usepackage{multirow}

\usepackage{hyperref}

\hypersetup{
    colorlinks=true,
    linkcolor=black,
    urlcolor=black,
    citecolor=black,
    breaklinks=true
}
\usepackage[left=1.62cm,right=1.62cm,top=1.9cm]{geometry}

\DeclareMathOperator*{\argmax}{argmax}
\DeclareMathOperator*{\argmin}{argmin}

\IEEEoverridecommandlockouts\IEEEpubid{\makebox[\columnwidth]{ 979-8-3503-1090-0/23/\$31.00~\copyright~2023 IEEE \hfill} \hspace{\columnsep}\makebox[\columnwidth]{ }}
\begin{document}

\title{Quantum Computing for MIMO
Beam Selection Problem: Model and Optical Experimental Solution
\thanks{*Corresponding authors: Yuhong Huang (huangyuhong@chinamobile.com), Wenxin Li (liwx@boseq.com), Kai Wen (wenk@boseq.com).}
}

\author{
\fontsize{9}{11}\selectfont
Yuhong Huang\textsuperscript{1}$^{*}$,
Wenxin Li\textsuperscript{2}$^{*}$,
Chengkang Pan\textsuperscript{1},
Shuai Hou\textsuperscript{1},
Xian Lu\textsuperscript{1},
Chunfeng Cui\textsuperscript{1} \\
Jingwei Wen\textsuperscript{3},
Jiaqi Xu\textsuperscript{4},
Chongyu Cao\textsuperscript{2},
Yin Ma\textsuperscript{2},
Hai Wei\textsuperscript{2},
Kai Wen\textsuperscript{2}$^{*}$\\
\fontsize{9}{11}\selectfont
\textsuperscript{1}\textit{China Mobile Research Institute, China}\\
\fontsize{9}{11}\selectfont
\textsuperscript{2}\textit{Beijing QBoson Quantum Technology Co., Ltd., Beijing 100015, China}\\
\fontsize{9}{11}\selectfont
\textsuperscript{3}\textit{China Mobile (Suzhou) Software Technology Company Limited, Suzhou 215163, China}\\
\fontsize{9}{11}\selectfont
\textsuperscript{4}\textit{Beijing University of Posts and Telecommunications}
}

\maketitle

\begin{abstract}
Massive multiple-input multiple-output (MIMO) has gained widespread popularity in recent years due to its ability to increase data rates, improve signal quality, and provide better coverage in challenging environments. In this paper, we investigate the MIMO beam selection (MBS) problem, which is proven to be NP-hard and computationally intractable. To deal with this problem, quantum computing that can provide faster and more efficient solutions to large-scale combinatorial optimization is considered. MBS is formulated in a quadratic unbounded binary optimization form and solved with Coherent Ising Machine (CIM) physical machine. We compare the performance of our solution with two classic heuristics, simulated annealing and Tabu search. The results demonstrate an average performance improvement by a factor of 261.23 and 20.6, respectively, which shows that CIM-based solution performs significantly better in terms of selecting the optimal subset of beams. This work shows great promise for practical 5G operation and promotes the application of quantum computing in solving computationally hard problems in communication.
\end{abstract}


\section{Introduction}

Massive multiple-input multiple-output (MIMO) is a revolutionary wireless transmission technology which has been introduced into 5G systems \cite{6736761}. By leveraging a high number of antennas and exploiting beamforming, it can offer multiple data streams simultaneously, allowing for higher throughput and better signal quality. This promises an improvement in cellular network coverage and capacity. 

However, due to the high mobility of users and inter-cell interference, the traditional relatively static beam setting can no longer meet the dynamic requirement of network coverage. Hence, MIMO beam selection (MBS) problem in 5G systems is guradully prominent. MBS refers to selecting a set of beams to maximize the performance of the networks under given constraints, such as improving signal quality and system throughput. Detailly, in MBS problem, the target geographic area is usually divided into square grids, and each beam is with a reference signal receive power (RSRP) value on the corresponding grid. The RSRP of one grid is defined as the maximum RSRP value of all beams on this grid. The MBS problem is to find a set of beams for every cell that maximizes the RSRP of each grids. The MBS problem is a hard combinatorial optimization problem, especially in 5G systems with a enormous number of cells and antennas. For example, when there are hundreds of beams in multiple cells, it is very difficult to find the best solution from billions of beam combinations. For combinatorial optimization problem \cite{schrijver2003combinatorial, li2022faster,li2023work,li2023performance}, there are some classical algorithms are proposed, such as greedy algorithms \cite{edmonds1971matroids,li2022submodular}, branch-and-bound \cite{wolsey2020integer}, and simulated annealing (SA) \cite{bertsimas1993simulated}. Greedy algorithms are simple and efficient, but they can get stuck in local optimum. Branch-and-bound algorithms can guarantee a global optimality, but can be computationally expensive. SA is a meta-heuristic optimization algorithm that gradually cools the temperature to encourage the optimization to converge to the global optimum while there is no guarantee to this.

Quantum computing is believed to have the potential to revolutionize the field of optimization by providing faster and more efficient solutions to large-scale combinatorial optimization problems. It can leverage the principles of quantum mechanics to perform multiple calculations simultaneously, allowing for the exploration of multiple solutions in parallel. There are several specific quantum computers in existence, one of which is the coherent ising machine (CIM) \cite{wang2013coherent, marandi2014network, mcmahon2016fully, inagaki2016coherent, honjo2021100, lu2023speed, lu2023recent}. CIM has been utilized in various applications, including solving problems such as compressed sensing \cite{aonishi2022l0}, polyomino puzzles \cite{takabatake2022solving} and job scheduling \cite{wen2023optical}. In this paper, we will explore the application of quantum computing in solving MBS problem.

\vspace{0.2cm}
\noindent \textbf{Related work:} There are lots of MBS schemes in massive MIMO systems. In \cite{DBLP:journals/tvt/JuLFHJ20}, the beam selection is formulated as a non-linear integer optimization problem. By deliberately transforming it to a geometric optimization problem and designing a rounding method, a suboptimal solution with low computational complexity can be achieved. In \cite{DBLP:journals/access/OrikumhiKJNK20}, a beam selection scheme is proposed subject to signal to interference plus noise ratio maximization, which is solved by Kuhn-Munkres assignment algorithm. However, most of the existing beam selection schemes are not practical due to the computational cost arising from iterative search or alternating optimization. Hence, the authors in \cite{reference} examine the complexity reduction of beam selection with incremental QR precoder and decremental QR precoder. The candidate beam size is decreased by using matrix perturbation theory to update QR decompositions. In \cite{DBLP:journals/access/HegdeS19}, the beam selection is modeled as a two-sided matching between users and beams with players preferences and becomes matching with externalities when considering the interdependences between users and beams. 


Several works have explored the utilization of CIM to address various optimization problems in the context of MIMO systems and error control decoding. For instance, \cite{singh2022ising} investigated the application of CIMs for MIMO Maximum Likelihood Detection (ML-MIMO). They introduced a regularized Ising approach, demonstrating the potential of CIM for achieving near-optimal MIMO detection. Prior to this, \cite{kim2019leveraging} focused on the application of Quantum Annealing (QA) to the MIMO detection problem, presenting the first system to apply QA to the computationally demanding ML MIMO wireless decoding problem. Additionally, \cite{kasi2023quantum} introduced a hybrid classical-quantum decoder design for Polar error correction codes, featuring the utilization of QA within a ML QA-based Polar decoder submodule.

\vspace{0.2cm}
\noindent \textbf{Our contributions:} The number of possible solutions in MBS problem scales exponentially with the number of cells and upper bound on the number of beams selected. This paper presents a novel approach to solve the MBS problem through the formulation of a quadratic unbounded binary optimization (QUBO) model. The CIM solver is utilized to solve the formulated QUBO model, which outperforms the classic heuristics. Our contribution are mainly in the following two terms:
\begin{itemize}
\item We develop an effective methodology to address the MBS problem, which can exploit the performance potential of MIMO cellular systems. Our proposed mathematical model in this paper is an elegant solution that effectively captures the underlying structure of the problem. We also further introduce a simplified model, which significantly reduces the number of bits required in the QUBO model, while is able to generate optimal solution. 

\item We validates the performance of CIM by conducting experiments on a physical CIM system. The experiments demonstrate that CIM is capable of generating optimal solutions for the MBS problem within an impressive millisecond-level time frame. This highlights the efficiency and effectiveness of CIM in finding solutions in real-world environments.
\end{itemize}

\section{Problem Formulation}\label{sec:formulation}
\begin{figure}[htbp]  \centerline{\includegraphics[width=0.35\textwidth]{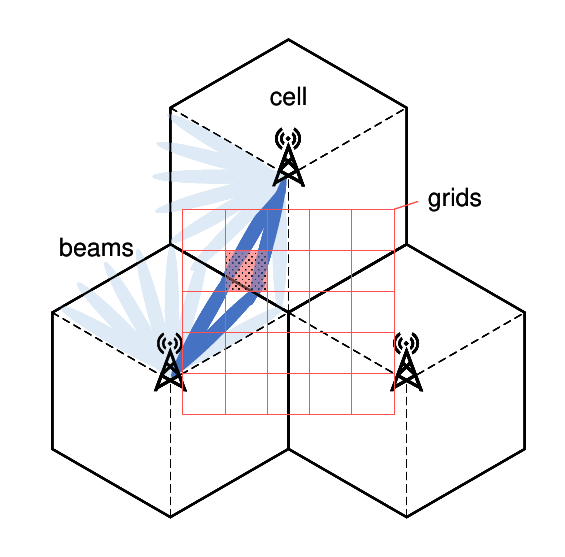}}
  \caption{A schematic illustration of the MBS problem.} 
  \label{base station} 
\end{figure}

As shown in Fig. \ref{base station}, in the MBS problem, the target coverage area is partitioned into small grids, each of which is covered by several cells. Each cell has a set of MIMO beams, and the MBS problem is to select a certain number of beams from each cell that maximize the number of grids satisfying certain constraints. A grid is deemed satisfactory if the maximum RSRP in the grid exceeds a given threshold, and the difference between the maximum RSRP and the second maximum RSRP in the grid exceeds a given value. The RSRP from a cell to a grid is determined by the maximum RSRP among all the beams. The reason for setting a threshold for the difference between the maximum and the second maximum signal strength is, in MIMO systems, there is mutual interference between beams. If the signal strengths of multiple beams are similar, it may cause signal interference and reduce the performance of the receiver. 
\vspace{-0.5cm} 

Let $n$ be the number of beams, $m$ be the number of grids. We use $v$ and $V_{i}$ to denote the number of cells and the set of cells associated with the $i$-th grid respectively. The RSRP varies at different grids, cells and beams. Let $s_{ijk}$ be the value of RSRP at the $i$-th grid, under the $j$-th cell and the $k$-th beam, and let $M$ be the maximum value of RSRP, i.e, $M = \max_{1\leq i \leq m, j \in V_i, 1\leq k \leq n}{s_{ijk}}$. The decision variables in MBS problem are $x_{jk}$s, where $x_{jk}=1$ if in the $j$-th cell, the $k$-th beam is selected. 

Now we are ready to characterize the RSRP at the $i$-th grid and $j$-th cell, i.e, $c_{ij}=\max_{1\leq k\leq n} s_{ijk}x_{jk}$, the maximum RSRP corresponds to the RSRP of the beam selected. 
\begin{align}
&c_{ij}\geq s_{ijk}x_{jk}, \forall j\in V_{i}, 1\leq k \leq n\label{cij1}\\
&c_{ij}\leq s_{ijk}x_{jk}+(1-d_{ijk})M, \forall 1\leq k \leq n \label{cij2}\\
&\sum_{k=1}^{n}d_{ijk}=1 \label{cij3}
\end{align}
We first note that $c_{ij}$ satisfies inequalities (\ref{cij1})-(\ref{cij2}) if $d_{ijk}\in \{0,1\} $ satisfies equation (\ref{cij3}), which can be verified based on the definition of $c_{ij}$. Moreover, we claim that (\ref{cij1})-(\ref{cij3}) implies that $c_{ij}=\max_{1\leq k\leq n} s_{ijk}x_{jk}$. Indeed, from (\ref{cij1}) we know that $c_{ij}\geq \max_{1\leq k\leq n} s_{ijk}x_{jk}$. Based on (\ref{cij2})-(\ref{cij3}), we can further obtain that $c_{ij}\leq \min_{k}{\{s_{ijk}x_{jk}+(1-d_{ijk})M\}}\leq\max_{1\leq k\leq n} s_{ijk}x_{jk}$,
where the last inequality holds when $d_{ijk}=1$ for $k=k^{*}=\argmax_{k}{s_{ijk}x_{jk}}$, and $d_{ijk}=0$ for $k\neq k^{*}$.

In a similar manner,  the maximum value of RSRP in the $i$-th grid can be characterized by the following constraints,
\begin{align}
&a_{i}\geq c_{ij}, \forall j\in V_{i} \label{ai1}\\       
&a_{i}\leq c_{ij}+(1-p_{ij})M, \forall j\in V_{i}\label{ai2}\\
&\sum_{j\in V_{i}}p_{ij}=1, p_{ij}\in \{0, 1\}\label{ai3}
\end{align}
Here $a_{i}=\max_{j\in V_{i}}{c_{ij}}$ is the unique solution to (\ref{ai1})-(\ref{ai3}). We emphasize that $p_{ij}=1$ only for $j=j^{*}=\argmax_{j}{c_{ij}}$, which will be used in the analysis that follows.

We next present the characterization of $b_{i}$, the second maximum RSRP in the $i$-th grid.
\begin{align}
b_{i}\geq c_{ij}-p_{ij}M, \forall j\in V_{i}   \label{bi1} \\ 
b_{i}\leq c_{ij}+(1-q_{ij})M, \forall j\in V_{i}   \label{bi3}\\ 
\sum_{j\in V_{i}}q_{ij}=2, q_{ij}\in \{0,1\}    \label{bi4}
\end{align} 
From (\ref{bi1}), we know that $b_{i}$ is no less than the maximum value of $c_{ij}-p_{ij}M$, i.e, $b_{i}\geq \max_{j}{c_{ij}-p_{ij}M}$. Recall that $p_{ij}=1$ when $c_{ij}$ is the maximum RSRP in grid $i$, hence $b_{i}$ is no less than the second maximum value of $c_{ij}$. From (\ref{bi3})-(\ref{bi4}), we obtain that $b_{i}\leq \min_{j}{c_{ij}+(1-q_{ij})M}$, where $\sum_{j\in V_{i}}q_{ij}=2$. It can be seen that $ \min_{j}{c_{ij}+(1-q_{ij})M}$ is no more than the second maximum value of $c_{ij}$. Hence we are able to conclude that $b_{i}$ satisfying (\ref{bi1})-(\ref{bi4}) must be the second maximum value of $c_{ij}$.

Now we introduce variable $z_{i} (1\leq i \leq m)$ to indicate whether the $i$th grid  satisfies the following two constraints:
\begin{itemize}
\item The maximum RSRP at cell $i$ is no less than $\delta_{1}$, i.e, $a_{i}\geq \delta_{1}$.
\item The difference between the maximum RSRP and second maximum RSRP is no less than threshold $\delta_{2}$, i.e, $a_{i}-b_{i}\geq \delta_{2}$.
\end{itemize}
Here, $\delta_{1}$ and $\delta_{2}$ are two predetermined thresholds specified in the constraints. Hence our objective function is $\max \sum_{i=1}^{m}{z_{i}}$,
where variables $z_{i}$s are subject to the following constraints:
\begin{align}
\delta_{1} - a_{i} &\leq M(1-z_{i})\label{zi1}\\
\delta_{2} - (a_{i} - b_{i}) &\leq M(1-z_{i})\label{zi2}
\end{align}
Inequalities (\ref{zi1}) and (\ref{zi2}) ensure that when the maximum RSRP is less than threshold $\delta_{1}$ or when the difference between the maximum RSRP and second maximum RSRP is less than $\delta_{2}$, indicator variable $z_{i}$ must be equal to $0$, otherwise $z_{i}$ will be equal to $1$, as our objective is to maximize the summation of $z_{i}$.

Finally, the number of beams selected should be less than or euqal to the predefined bound, i.e., 
\begin{align}\label{cardcons}
\sum_{k=1}^{n}x_{jk} &\leq r, \;\forall 1\leq j \leq v 
\end{align}

Due to space limitation, in \cite{mimo} we show that The MBS problem defined by (\ref{cij1})-(\ref{cardcons}) is NP-hard. 

Inequalities (\ref{cij1})-(\ref{cij2}), (\ref{ai1})-(\ref{ai2}), (\ref{bi1}), (\ref{bi3}) and (\ref{zi1})-(\ref{zi2}) can be transformed into equality by introducing slack variables, which can be expressed with binary expansion as following:
\begin{align}\label{slack}
\mathrm{slack} = \sum_{t=0}^{\ell} 2^{t}\cdot \mathrm{slack}_{t}
\end{align}
In (\ref{slack}), $\mathrm{slack}_{t}$ takes values of zero or one, $\ell$ is related to the range of slack variables. Here we let $\ell = \lceil\log_{2}M\rceil$, as all the slack variables are no more than $M$.

We next replace the aforementioned equality constrained maximization problem by an unconstrained minimization whose optimal solution is identical to that of the original constrained maximization problem. The unconstrained problem is formed by adding a penalty function to the objective function. The penalty function consists of a parameter $\lambda$ multiplies a measure of violation of the constraints, i.e., the square of the equality constraints. 

The details of unconstrained minimization problem is presented in \cite{mimo}. $a_{i}$, $b_{i}$ and $c_{ij}$ are integer variables, and can be expressed using binary representation: $a_{i} = \sum_{t=1}^{\ell} 2^{t} \cdot a_{it}$, $b_{i} = \sum_{t=1}^{\ell} 2^{t} \cdot b_{it}$, and $c_{ij} = \sum_{t=1}^{\ell} 2^{t} \cdot c_{ijt}$.

Now the optimization problem has QUBO form. While the computational complexity of quadratic optimization grows exponentially with the number of beams, leveraging a quantum computer to implement a QUBO problem has the potential to outperform classical solvers in terms of speed. In QUBO problem, we have a set of binary vectors of a fixed length $n$, and we are given a real-valued upper triangular matrix $Q\in \mathbb{R}^{n\times n}$. Function $f_{Q}:\{0,1\}^{n}\rightarrow \mathbb{R}$ assigns a value to each binary vector through $f_{Q}(\mathbf{x}) = \mathbf{x}^{\mathrm{T}}Q\mathbf{x}=\sum_{i\leq j}Q_{ij}x_{i}x_{j}$. 
The objective of the QUBO function is to find a binary vector $\mathbf{x}^{*}$ that minimize $f_{Q}$, namely, $\mathbf{x}^{*} =\argmin_{\mathbf{x}} f_{Q}(\mathbf{x})$.
QUBO is highly related to the Ising model, whose Hamiltonian function is given as, $H(\sigma) = -\sum_{i,j}J_{ij}\sigma_{i}\sigma_{j}-\sum_{i}{h_{i}\sigma_{i}}$,  
where $\sigma_{i}$ is the spin variables taking value from $\{-1, +1\}$. $J_{ij}$ and $h_{i}$ are quadratic and linear coefficients respectively. Utilizing the identity $\sigma\rightarrow 2x-1$, we can obtain the equivalent QUBO problem.

The optical CIM system is capable of implementing the Ising Hamiltonian with adjustable parameters, which allows for the transformation of a QUBO problem to be solved. By utilizing a controllable quantum phase transition process, the optimal solution can be obtained. To demonstrate this, in Section \ref{experiment} we have conducted experiments involving the quantum optimization algorithm for MBS problem, in comparison to some classical algorithms.

\section{An Alternative Simplified Model With Post-Processing}\label{sec:simplifiedmodel}

To reduce the number of bits in the QUBO model, here we investigate a simplified model in which only constraint that the maximum RSRP is no less than threshold $\delta_{1}$, i.e., $a_{i}\geq \delta_{1}$ is considered. We obtain the best $100$ solutions returned by CIM, from which we select the best feasible solution of the origin model defined in Section \ref{sec:formulation}.

Let $\bar{s}_{ijk}=1$ if $s_{ijk}\geq \delta_{1}$, otherwise $\bar{s}_{ijk}=0$. Similar as the formulation in Section \ref{sec:formulation}, the simplified model can be expressed as,
\begin{align}
\max &\sum_{i=1}^{m}z_{i}\\
\operatorname{s.t.} \ &z_{i}\leq \sum_{j\in V_{i}}\sum_{k=1}^{n} x_{jk}\bar{s}_{ijk}, \forall 1\leq i \leq m\label{model21}\\
&\sum_{k=1}^{n}x_{jk}\leq r, \forall 1\leq j\leq v\label{model22}
\end{align}
We argue that inequality (\ref{model21}) ensures that $z_{i}=1$ if and only if the constraint that maximum RSRP is no less than $\delta_{1}$ is satisfied. Note that if $x_{jk}\bar{s}_{ijk}=0$ for $\forall j\in V_{i}, 1\leq k\leq n$, i.e., there exists no RSRP being selected that is no less than the threshold $\delta_{1}$, $z_{i}$ must be equal to $0$, according to (\ref{model21}). Otherwise if there exists $j$ and $k$ such that $x_{jk}\bar{s}_{ijk}=1$,  $z_{i}$ will be equal to $1$.

In the following we present the QUBO form of the simplified model, by introducing slack variable for constraint (\ref{model21}) and (\ref{model22}):
\begin{align}\label{model2qubo}
 \min -\sum_{i=1}^{m}{z_{i}} &+ \lambda \cdot \Big\{\sum_{i=1}^{m}\Big(z_{i}+\mathrm{slack}_{1,i}-\sum_{j\in V_{i}}\sum_{k=1}^{n} x_{jk}\bar{s}_{ijk}\Big)^2\notag\\
&+\sum_{j=1}^{v}\Big(\sum_{k=1}^{n}x_{jk}+\mathrm{slack}_{2,j}- r\Big)^2\Big\}.   
\end{align}

It can be verified that there are $(m+nv+m\cdot \lceil\log_{2}(nv)\rceil+v\cdot\lceil\log_{2}r\rceil)$ binary variables in this simplified model.

\section{Experimental Results---Comparision between CIM physical machine and classic heuristics}\label{experiment}

In this section, we aim to to evaluate the feasibility of using CIM as a solver for the MBS problem with respect to solution quality and problem size that can fit the CIM. The experimental data is gathered by measuring signal strength within a contiguous region in Ji'an City, China. The dataset consists of signal data from $4857$ grids, $217$ cells, and $148$ beams, and contains $1048575$ data records.

\begin{figure}[htbp]
\centerline{\includegraphics[width=0.4\textwidth]{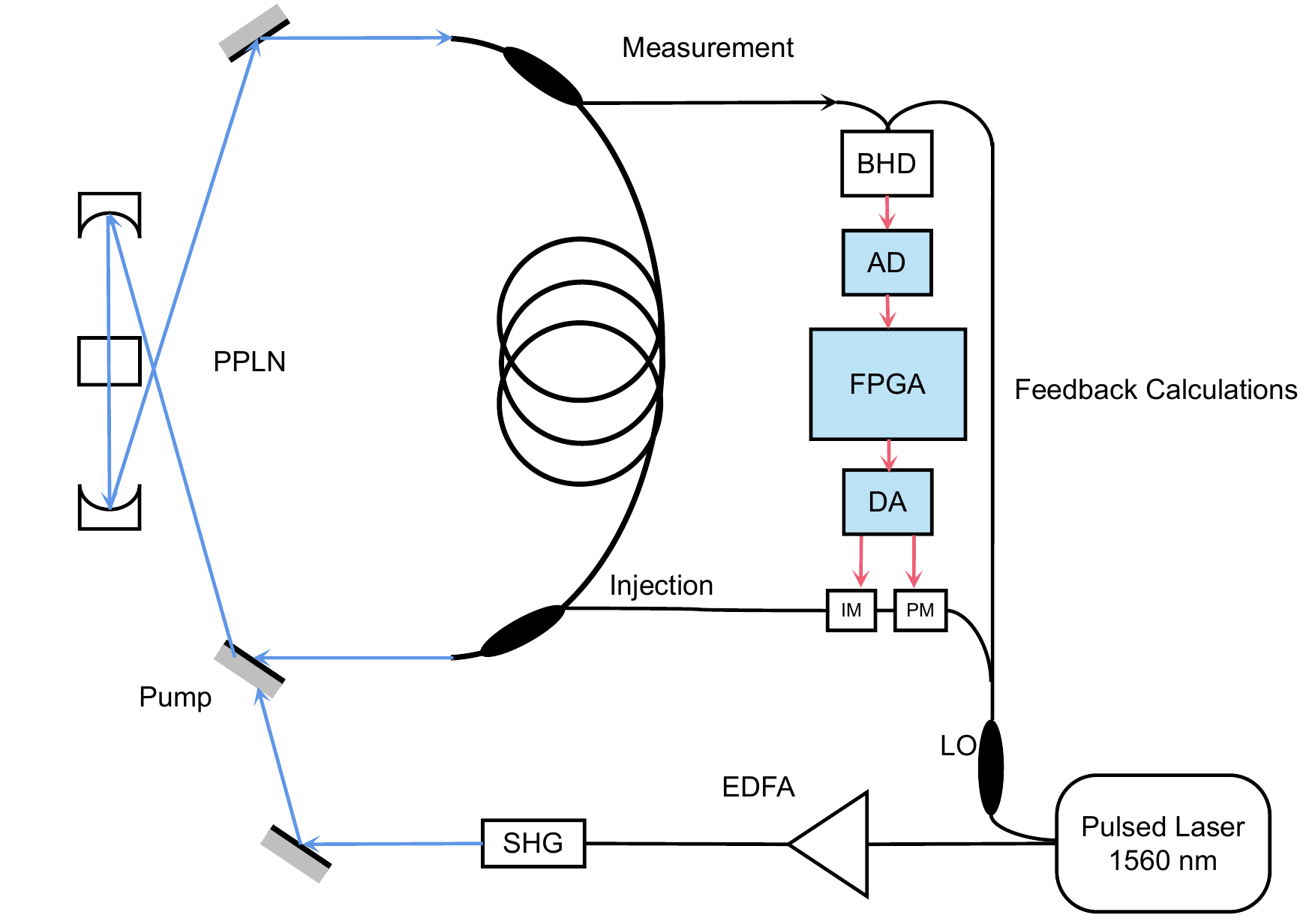}}
  \caption{Illustration of CIM.}
  \label{fig:example}
\end{figure}

As shown in Fig. \ref{fig:example}, CIM operates differently from traditional computers that rely on semiconductor integrated chips. Instead, it employs laser pulses in optical fibers as the basic units of computation, referred to as qubits. The initial research was centered around the idea of the injection synchronous laser Ising machine \cite{utsunomiya2011mapping}. As the number of coupled lasers grows in proportion to the square of qubits, based on the Degenerate Optical Parametric Oscillator (DOPO), an improved scheme where nonlinear optical crystals are used was proposed. Two types of DOPO-based approaches have been developed, which are the optical delay lines CIM \cite{marandi2014network} and the measurement feedback CIM \cite{haribara2015coherent}. However, in the first scheme, the overhead and the requirement of precise control is not affordable. This paper is based on the measurement feedback CIM developed by Beijing Qbosen Quantum Technology
Co., Ltd~\cite{qboson}.

In this experiment, we consider the scenario where both the beam and cell numbers are fixed at 5, while the grid size is varied from 5 to 10. All six settings involve bit numbers less than 100, and were chosen to investigate the performance of the CIM physical machine under different grid sizes while controlling for other variables. To evaluate the CIM physical machine, we use the Max-Cut problem, which is a well-known NP-hard problem. It can be shown that finding the lowest energy state of the Ising model without a magnetic field can be reformulated as Max-Cut Problem \cite{honjo2021100}. 

\begin{figure*}[h]
  \centering
  \begin{subfigure}[b]{0.325\textwidth}
    \includegraphics[width=\textwidth]{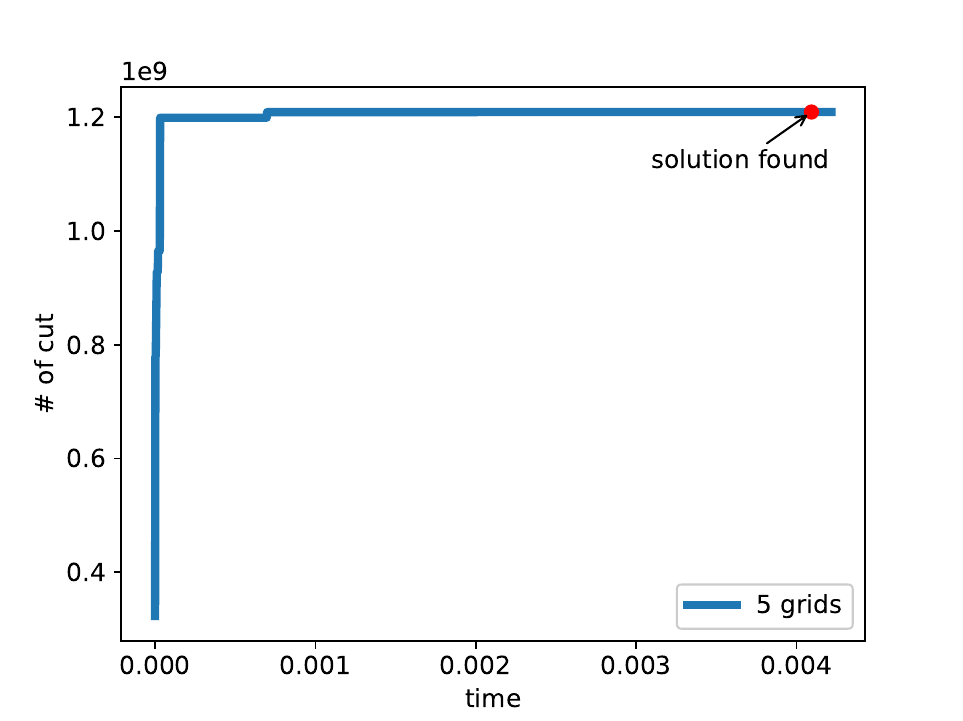}
  \end{subfigure}
  \hfill
  \begin{subfigure}[b]{0.325\textwidth}
    \includegraphics[width=\textwidth]{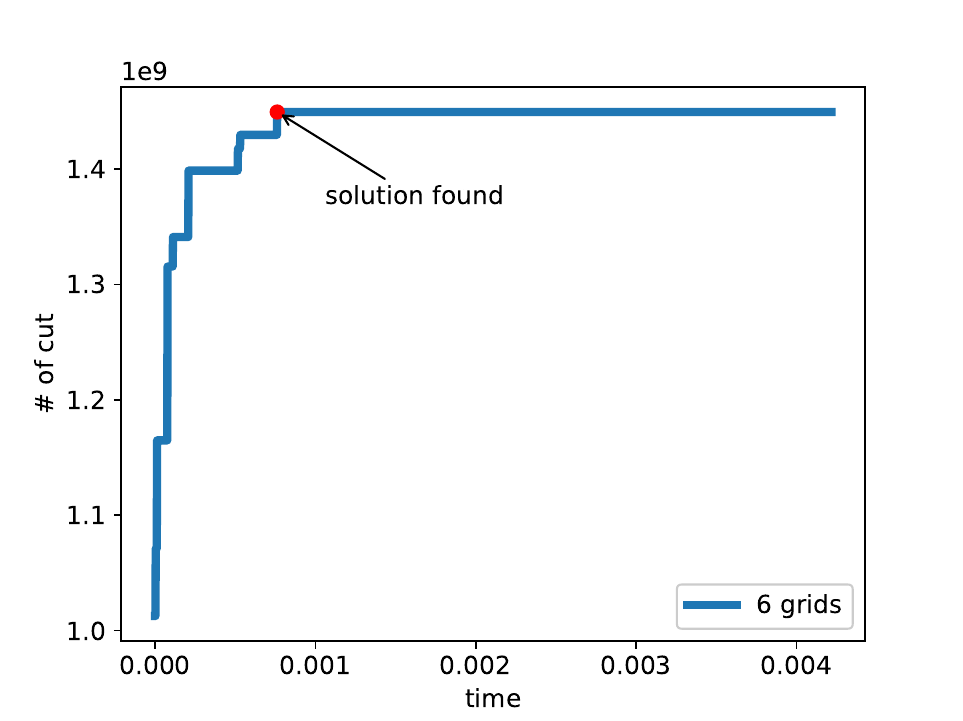}
  \end{subfigure}
  \hfill
  \begin{subfigure}[b]{0.325\textwidth}
    \includegraphics[width=\textwidth]{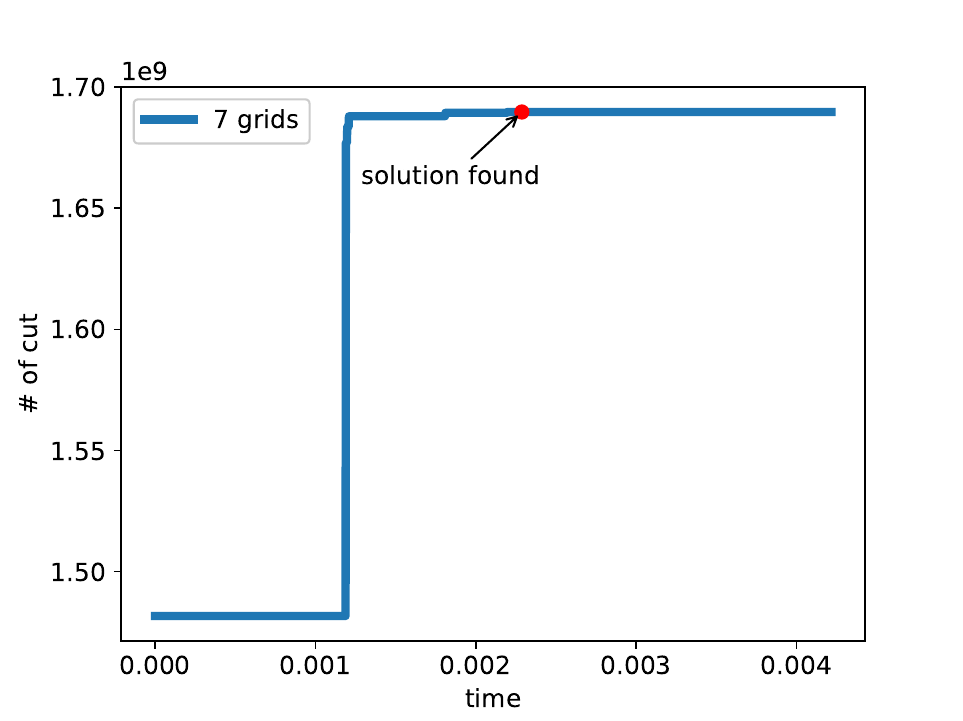}
  \end{subfigure}
  \hfill
  \begin{subfigure}[b]{0.325\textwidth}
    \includegraphics[width=\textwidth]{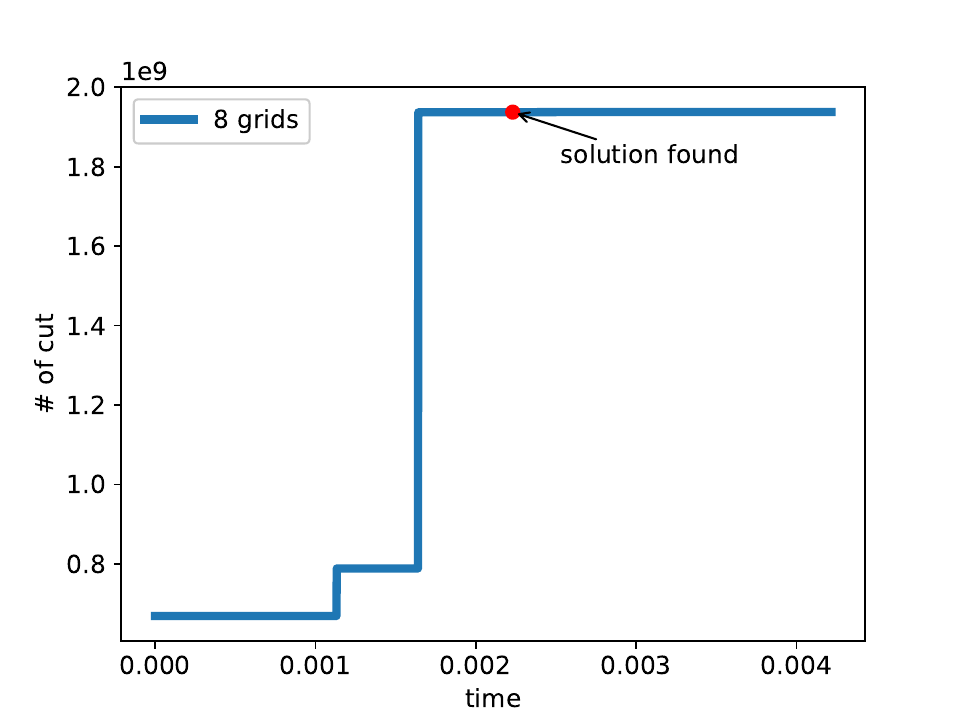}
  \end{subfigure}
  \hfill
  \begin{subfigure}[b]{0.325\textwidth}
    \includegraphics[width=\textwidth]{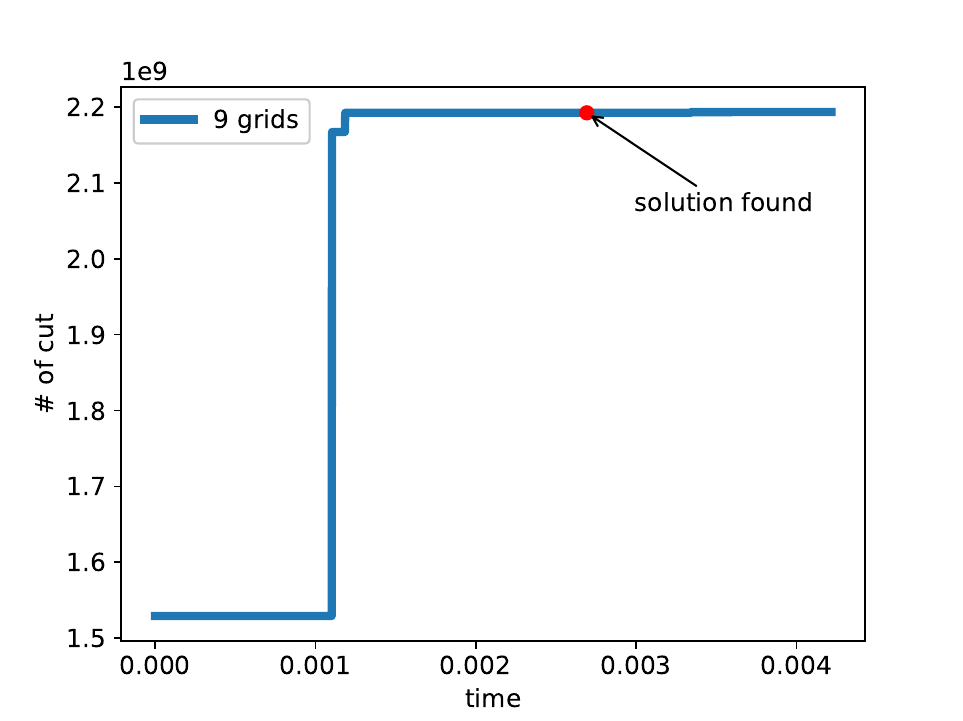}
  \end{subfigure}
  \hfill
  \begin{subfigure}[b]{0.325\textwidth}
    \includegraphics[width=\textwidth]{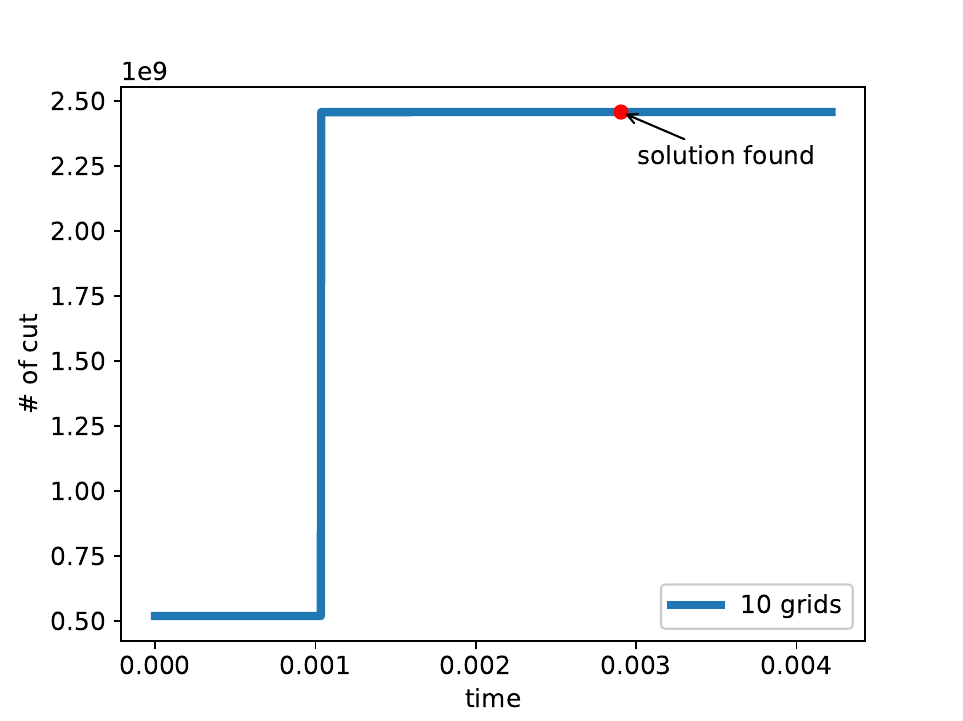}
  \end{subfigure}
  \caption{Evolution process of the cut value.}
  \label{fig:cut value}
\end{figure*}

In Fig. \ref{fig:cut value}, we have plotted the cut value obtained by the CIM as a function of computation time. The data points display the cut value that were assessed at the intermediate stages of the CIM. The temporal spacing between every two data points is 2.11 $\mu$s. Indeed, there were $211$ oscillating pulses in the fiber loop with a time interval of $10$ ns between every two pulses. Consequently, the transmission time of optical pulses in the loop is 2.11 $\mu$s. We can see that the cut value increases over time, and as the power of the pump light is gradually increased towards the threshold for oscillation, a phase transition occurs.

We seek feasible solutions that optimize the objective function in the vicinity of the lowest energy. The optimal solution found is denoted with a red dot, as depicted in Fig. \ref{fig:cut value}. Among the six scenarios, the CIM machine employed $1940$, $361$, $1084$, $1057$, $1276$, and $1377$ circulations, respectively, to find the optimal solution, with corresponding runtimes of 4.096ms, 0.764ms, 2.289ms, 2.232ms, 2.694ms and 2.908ms.

\begin{figure}[htbp]
  \centerline{\includegraphics[width=0.3\textwidth]{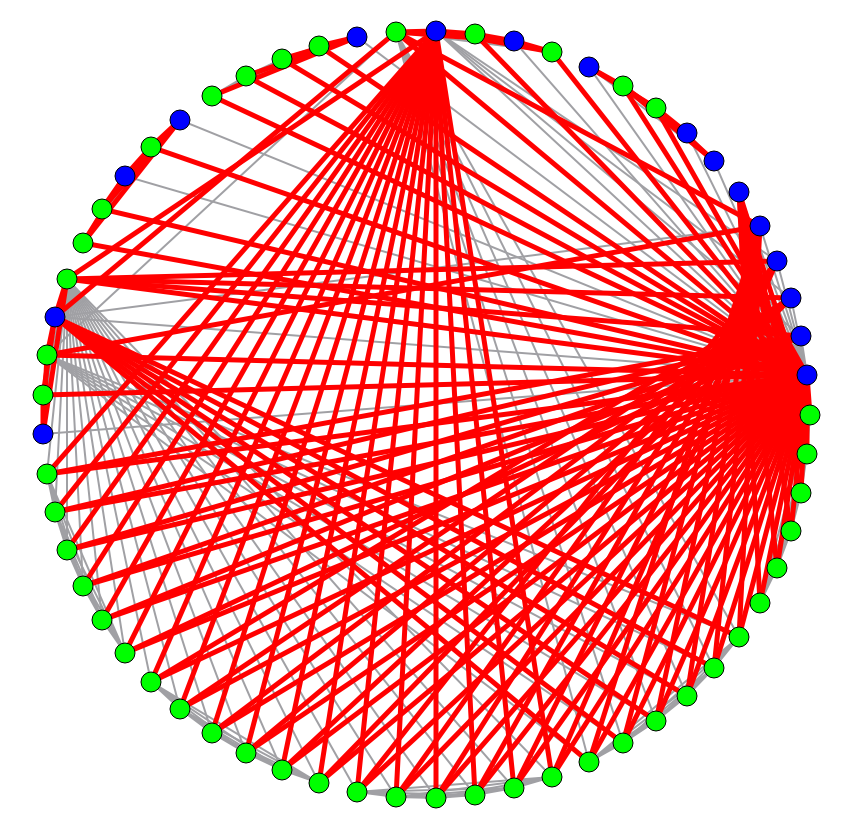}}
  \caption{Result found by CIM physical machine for $m=5$.}
  \label{fig:cut}
\end{figure}

\begin{table*}[!htbp]
\resizebox{\linewidth}{!}{
\begin{tabular}{|l|l|l|l|l|l|l|l|l|l|l|}
\hline
      \# of grids & \multirow{2}{*}{ \# of bits} & \multicolumn{3}{l|}{CIM Physical Machine} & \multicolumn{2}{l|}{Optimal Solution} & \multicolumn{2}{l|}{SA} & \multicolumn{2}{l|}{Tabu} \\
\cline{3-11}
       & & Time & Value & Hamiltonian &  Value & Hamiltonian & Time & Value & Time & Value \\
\hline
m = 5 & 61& 4.096ms&5 &-1636500253 & 5& -1636753751.5 & 134ms & 2.07& 13.7ms&1.8 \\
m = 6 & 68& 0.764ms&6 &-1961998250 & 6&-1962503752 & 147ms & 1.55& 14.3ms&2.6 \\
m = 7 & 75&2.289ms &7 & -2286992748&7 & -2288253752.5&131ms & 1.87&17.3ms &2.94 \\
m = 8 & 82& 2.232ms&7 & -2619998246& 7& -2621753753& 133ms & 2.28& 16.7ms&3.15 \\
m = 9 & 89&2.694ms & 7& -2899621246&7 & -2963128752.5& 139ms & 2 & 20.2ms&3.04 \\
m = 10 &96&2.908ms & 7& -3310865745& 7&-3312378751 & 146ms & 2.12 & 22ms&2.7 \\
\hline
\end{tabular}}
\caption{Comparision of CIM physical machine, SA and Tabu search in objective value and running time.}
\label{physical_machine_table}
\end{table*}

As an illustrative example, Fig. \ref{fig:cut} shows the solution obtained by the CIM physical machine when there are $5$ grids, where blue nodes represent $+1$ and green dots represent $-1$ for spins. We can observe that the corresponding graph has many edges between the points and is nearly fully connected, which implies that our MBS problem is highly sophisticated.

Table \ref{physical_machine_table} presents a comprehensive performance comparison between CIM physical machine, SA and Tabu search. Regarding the objective value achieved by the algorithms, CIM Physical Machine consistently achieved the best values of the objective function for all cases. In addition, the Hamiltonian of the solution found by the CIM physical machine is very close to that of the optimal solution, which confirms the effectiveness of our approach to simplifying the model and searching for the solution near the lowest energy value. In order to evaluate the performance of the SA and Tabu search algorithms, we conduct 100 repetitions and calculate the mean time to find a feasible solution, as well as the mean objective function value corresponding to the feasible solution found.

\begin{figure}[htbp]
\centerline{\includegraphics[width=0.44\textwidth]{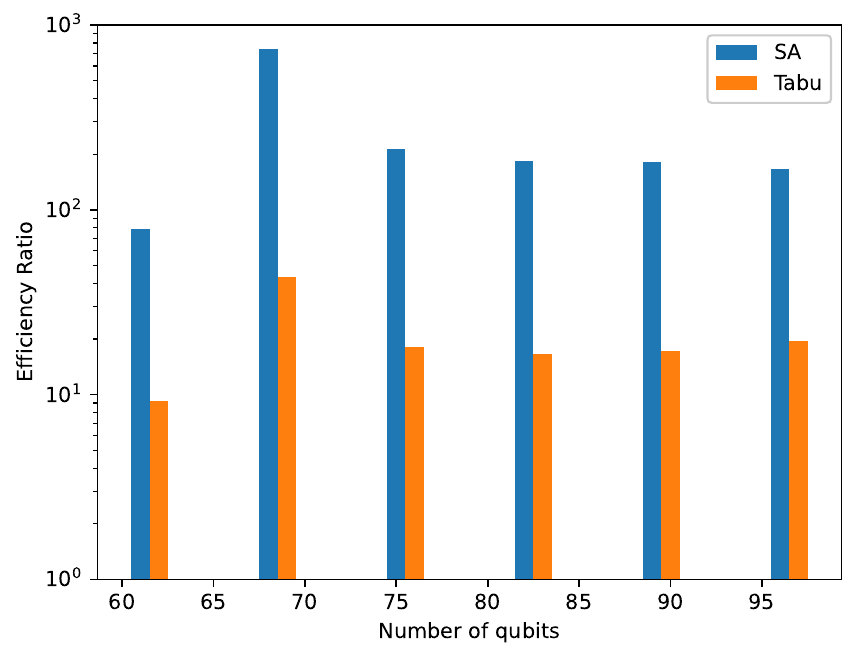}}
  \caption{Variation of efficiency ratio with number of bits.}
  \label{fig:efficiency ratio}
\end{figure}

Let $t_{\mathrm{CIM/sa/tabu}}$, $f_{\mathrm{CIM/sa/tabu}}$ be the running time and objective value of the corresponding algorithm, respectively. We define the following \emph{efficiency ratio} to evaluate the ability of CIM,
\begin{align}
\gamma_{\mathrm{sa}/\mathrm{tabu}} = \frac{f_{\mathrm{CIM}}/t_{\mathrm{CIM}}}{f_{\mathrm{sa}/\mathrm{tabu}}/t_{\mathrm{sa}/\mathrm{tabu}}}.
\end{align}

The efficiency ratio is based on two components: the time-saving effect and accuracy improvement. The time-saving effect measures the amount of time saved by using CIM compared to traditional methods, while the accuracy improvement measures the extent to which the algorithm's output matches the correct solution, compared to SA and Tabu. A greater efficiency ratio implies a better performance of CIM. It can be seen that CIM achieves tens of times of performance improvement to Tabu and hundreds of times of improvement to Tabu search. Indeed, comparing to SA and Tabu, CIM obtains an average efficiency ratio of $261.23$ and $20.66$ respectively.

\section{Conclusion}
In this paper, we propose a new approach based on a QUBO  model to address the MBS problem, which is a key issue of 5G system. Besides the carefully designed model, our contributions also include the successful application of the CIM simulator and physical machine to solve the MBS problem. The empirical results show that the CIM Solver outperforms classic heuristics in terms of accuracy and speed, providing  tens to hundreds of times performance improvement. We believe that the proposed solution has great potential for practical 5G network operation. The thoughtful design of our model highlight our contributions to the field, paving the way for further advancements in this area of research.

\bibliographystyle{ieeetr}
\bibliography{reference}

\end{document}